\documentclass[12pt]{article}
\usepackage{times}
\usepackage{natbib}
\usepackage{graphics,graphicx}
\usepackage{geometry}
\usepackage[utf8]{inputenc}
\usepackage{enumitem,amssymb}
\usepackage{ragged2e}
\newlist{thematic}{itemize}{8}
\setlist[thematic]{label=$\square$}
\usepackage{pifont}

\begin{document}
\raggedright
\huge
Astro2020 Science White Paper \linebreak

Studying Magnetic Fields in Star Formation
and the Turbulent Interstellar Medium\linebreak
\normalsize

\noindent \textbf{Thematic Areas:} \hspace*{60pt} $\square$ Planetary Systems \hspace*{10pt} $\blacksquare$ Star and Planet Formation \hspace*{20pt}\linebreak
$\square$ Formation and Evolution of Compact Objects \hspace*{31pt} $\square$ Cosmology and Fundamental Physics \linebreak
  $\square$  Stars and Stellar Evolution \hspace*{1pt} $\square$ Resolved Stellar Populations and their Environments \hspace*{40pt} \linebreak
  $\square$    Galaxy Evolution   \hspace*{45pt} $\square$             Multi-Messenger Astronomy and Astrophysics \hspace*{65pt} \linebreak
  
\textbf{Principal Author:}

Name: Laura Fissel
 \linebreak						
Institution: NRAO
 \linebreak
Email: lfissel@nrao.edu
 \linebreak
Phone: (434)244-6821
 \linebreak
 
\textbf{Co-authors:} Charles L. H. Hull (NAOJ/ALMA), Susan E. Clark (Institute for Advanced Study), David T. Chuss (Villanova), Philippe Andr\'{e} (CEA Saclay), Fran\c{c}ois Boulanger (Universite Paris Sud/IAS), C. Darren Dowell (JPL/Caltech), Edith Falgarone (ENS), Brandon Hensley (Princeton), A. Lazarian (U. Wisconsin, Madison), Giles Novak (Northwestern), Ian Stephens (Harvard CfA), Siyao Xu (U. Wisconsin, Madison) 
  \linebreak

\textbf{Abstract:}
Understanding the physics of how stars form is a highly-prioritized goal of modern Astrophysics, in part because star formation is linked to both galactic dynamics on large scales and to the formation of planets on small scales.  It is well-known that stars form from the gravitational collapse of molecular clouds, which are in turn formed out of the turbulent interstellar medium. Star formation is highly inefficient, with one of the likely culprits being the regulation against gravitational collapse provided by magnetic fields.  Measurement of the polarized emission from interstellar dust grains, which are partially aligned with the magnetic field, provides a key tool for understanding the role these fields play in the star formation process. 

Over the past decade, much progress has been made by the most recent generation of polarimeters operating over a range of wavelengths (from the far-infrared through the millimeter part of the spectrum) and over a range of angular resolutions (from less than an arcsecond through fractions of a degree). Future developments in instrument sensitivity for ground-based, airborne, and space-borne polarimeters operating over range of spatial scales are critical for enabling revolutionary steps forward in our understanding of the magnetized turbulence from which stars are formed.

\pagebreak

\justify

\section{Introduction}
\label{sec:intro}

The magnetic field is a key regulator of star formation at all spatial scales \citep{Elmegreen2004,Crutcher2012}.  
Magnetic fields shape turbulence over a broad 
range of length scales in the diffuse interstellar medium (ISM)
and are a critical factor in the formation and evolution of molecular clouds as well as protostellar cores, envelopes, disks, outflows, and jets.   The efficiency of the conversion of molecular gas to stars during the star-formation process is notoriously low, due to regulation from a variety of sources including turbulent gas motions, magnetic fields, and feedback from young stars \citep{McKee2007,Federrath2015}.  In particular, magnetic fields play an important role in slowing the process of star formation by inhibiting movement of gas in the direction perpendicular to the field lines.  

Over the last two decades a vast amount of work has focused on characterizing magnetic fields and turbulence in the ISM from galactic to cloud to protostellar core scales, using interferometers as well as ground-, balloon-, and space-based single-dish instruments. By measuring polarized thermal emission from dust grains aligned with respect to the magnetic field, we can map the magnetic field morphology in star forming regions \citep{Jones1984, LazarianHoang2007,Andersson2015}. The \textit{Planck} satellite recently mapped dust polarization across the entire Milky Way at a best resolution of 5$^{\prime}$~\citep{PlanckXIX}. However, \textit{Planck} lacks the sensitivity needed to study magnetic fields and turbulence in detail in all but the very nearest molecular clouds, and cannot resolve dense filamentary structures and cores in any cloud.  In contrast, ground-based polarimeters have produced high-resolution maps of dense sub-regions within molecular clouds; however, such instruments are sensitivity-limited as a result of observing through the atmosphere, and can therefore typically only map very small ($<\,$0.01\,deg$^2$), bright regions within molecular clouds (e.g.~\citealt{Dotson2010,Matthews2009}).

 High-sensitivity observations of a large number (of order thousands) of clouds, in addition to extremely well resolved maps of the well known set of nearby molecular clouds, will allow us to answer key questions: How do magnetic fields affect the formation of molecular clouds from the diffuse interstellar medium?  And do magnetic fields influence the formation and evolution of gravitationally unstable cloud sub-structures, and consequently, the efficiency with which stars form?

\section{The diffuse magnetized medium}

Molecular clouds condense out of the diffuse ISM, and so the process of star formation is regulated in its earliest phases by the flow of material into the molecular phase. Magnetic fields influence structure formation directly, especially through magnetic pressure, and via their influence on the acceleration and propagation of cosmic rays \citep{Strong:2007} and the transfer of heat and matter between the hot, warm, and cold phases of the ISM \citep{Draine2011}. In addition to being the fuel for star formation, the diffuse ISM is also the medium that responds to supernova explosions and other large-scale injections of energy. Despite its importance, a comprehensive understanding of the magnetized diffuse ISM is challenging because of its diverse composition, its sheer expanse, and the multi-scale nature of the physics that shapes it. 

Recent high-resolution observations of the diffuse ISM have revealed a wealth of complex structure, deeply influenced by magnetized turbulence and the structure of the ambient magnetic field. The structure of the cold neutral medium (CNM) in particular is highly anisotropic, largely organized into filamentary structures that are aligned with the magnetic field \citep{Clark2014}. A preferred alignment between ISM structures and the magnetic field traced by dust polarization has recently been measured in both neutral hydrogen \citep{Clark2015, Kalberla:2016} and dust emission \citep{PlanckXXXII, Malinen2016}. 
Sensitive observations of polarized dust emission toward diffuse regions of the ISM are crucial for understanding the dynamical importance of the magnetic field for structure formation at a given scale.

The diffuse ISM is an ideal laboratory for studying magnetohydrodynamic (MHD) turbulence. Turbulence correlates matter and energy over a range of scales, traditionally parameterized as a cascade of energy across the wavenumber spectrum \citep{Elmegreen2004}. Studying turbulence necessitates high dynamic range observations in order to have robust statistical samples on many different scales, and dust polarization provides one of the few avenues for high-dynamic-range magnetic field observations. One statistical measure is the cross-correlation between dust temperature ($T$) and the dust polarization field, decomposed into the rotationally invariant $E$- and $B$-modes. \textit{Planck} measured these quantities, with several unexpected results. There is a positive $TE$ correlation in the diffuse ISM, as well as a non-unity ratio of $E$-mode to $B$-mode polarization ($EE/BB \sim 2$). 
The $TE$ and $EE/BB$ correlations are likely due to magnetically induced anisotropy in the dust distribution \citep{Clark2015, Planck:XXXVIII}, and to MHD turbulence parameters like the Alfv\'en Mach number \citep[e.g.,][]{Caldwell:2017, Kandel:2017}. 
Perhaps even more surprising, \textit{Planck} measured a nonzero $TB$ correlation across a range of scales in the diffuse ISM \citep{Planck2018:XI}. The nonzero $TB$ measurement could be a signature of a large-scale helical magnetic field \citep[e.g.][]{Bracco:2019}.

Detailed studies of the diffuse ISM are highly complementary to the search for inflationary $B$-mode polarization in the cosmic microwave background (CMB), as both science cases require sensitive polarimetric observations of low-column-density regions of sky. In addition, a better understanding of the magnetized ISM allows better removal of the polarized foreground for next-generation CMB polarization experiments. This is critical for the current generation of ground and balloon-borne CMB experiments \citep{ACT2016,EBEX2018,PIPER2016,SPIDER2018,BICEPKECK2018,Simons2018,CLASS2016}, and will be even more important for future experiments that will have increased sensitivity \citep{S42017,PICO2019,Litebird2018}.  

\section{The role of magnetic fields in molecular clouds} 

Current observations suggest that the outer envelopes of molecular clouds can be supported against gravity by magnetic fields at $\sim$\,parsec scales, but that gravity dominates at the $\lesssim$\,0.1\,pc scale of dense cores, and so these dense structures can collapse to form stars \citep{Crutcher2010}. Furthermore, \textit{Planck} observations of the ten nearest (mostly low-mass) molecular clouds show a change in alignment of cloud structure from parallel to perpendicular to the magnetic field with increasing column density \citep{PlanckXXXII, PlanckXXXV}; this effect can only be reproduced in simulations if the cloud-scale magnetic energy density is equal to or larger than the energy density of turbulence \citep{Soler2013,CYChen2016,Mocz2018}.  More observations are needed to determine whether this ``strong field'' case is true for giant molecular clouds, to measure the physical scale at which molecular clouds make the transition from being magnetically dominated to turbulence- and/or gravity-dominated, and to determine whether the cloud magnetic field strength affects star formation efficiency within molecular clouds.

\begin{figure}
    \centering
    \begin{center}
    \includegraphics[width=\textwidth]{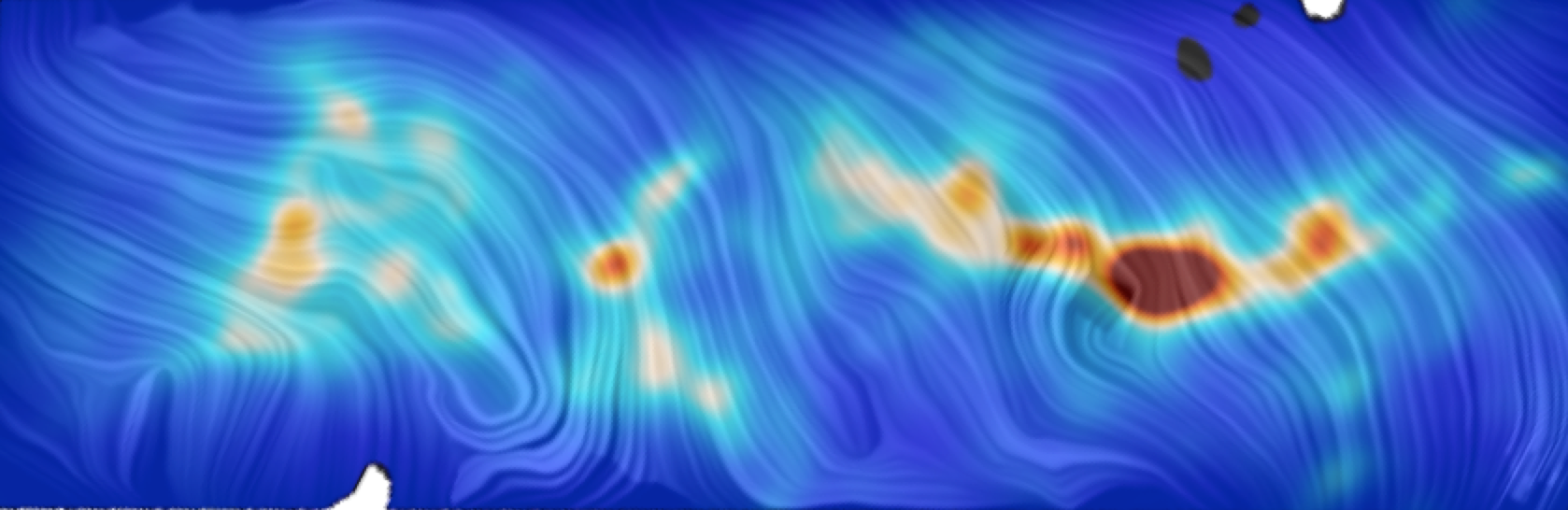}
    \end{center}
    \caption{
    \textit{
    500\,$\mu$m image of the Vela C molecular cloud from the balloon-borne BLASTPol instrument; the image has a resolution of 2.5$'$, or $\sim$\,500\,pc at the distance to Vela C \citep{Fissel2016}.  Next-generation space-based single-dish missions will enable even higher resolution and higher sensitivity images of nearby molecular clouds.
    }}
    \label{fig:vela}
\end{figure}

Polarimeters proposed as part of next-generation space observatories, such as OST, PICO, and SPICA, will have better sensitivity to low surface brightness dust and a higher spatial dynamic range and  could build on the progress made by \textit{Planck}, SOFIA, and BLASTPol in the area of detailed studies of nearby molecular clouds (such as the BLASTPol observations of Vela C: see Figure \ref{fig:vela}, from \citealt{Fissel2016}).  Full-sky coverage polarization maps at a resolution of 1$^{\prime}$, would be especially helpful since all maps would have better than 1\,pc resolution out to a distance of 3.4\,kpc, yielding detailed maps of magnetic fields in over 2,000 molecular clouds with 1000s to 100,000s of independent measurements per cloud.  These observations will complement other methods for probing magnetic fields in molecular clouds such as Zeeman splitting of molecular lines \citep{Crutcher2010}, analysis of velocity structure \citep{LazarianYuen2018b}, and optical/near-IR polarization due to selective extinction of background starlight \citep{Marchwinski2012}.

Observing a large sample of clouds is crucial because measurements of polarization angle are sensitive to only the magnetic field orientation projected on the plane of the sky.  Polarization maps will look very different for molecular clouds for different viewing angles; for young clouds versus evolved clouds; and for low-mass clouds compared with high-mass star forming regions.  
With future balloon and space polarimeters providing detailed maps of magnetic fields in hundreds of molecular clouds it will be possible to compare cloud magnetization to the efficiency of star formation, measured from near and far-IR observations of dense cores and protostars with {\em Herschel}, {\em Spitzer}, and {\em JWST}. With observations of thousands of molecular clouds it should be possible to determine the role of magnetic fields in star formation as a function of age and mass.

As a first step toward this goal, balloon-borne sub-millimeter polarimeters such as the upcoming BLAST-TNG telescope \cite{Galitzki2014} and the PILOT experiment \citep{Bernard2016}, which operate above $>$99.5$\%$ of the atmosphere, will be able to map dozens of molecular clouds with $<$\,1$^{\prime}$~resolution. However, polarization capabilities on next-generation far-IR satellites, which can be cooled to $<$10\,K and therefore will have orders of magnitude improvements in sensitivity, will enable much more detailed studies of the nearest molecular clouds. With 10$^{\prime \prime}$~FWHM resolution (15$\times$~higher than the BLASTPol map of Vela\,C shown in Figure \ref{fig:vela}), such a telescope could measure the cloud magnetization and MHD turbulence over a wide range of density scales, from the diffuse cloud envelope, to the scale of dense, $<$0.1\,pc width filaments and cores. 

\section{Do magnetic fields affect the collapse of cores and filaments?}

Filamentary structures within molecular clouds appear to contain most of the dense, gravitationally unstable gas and are observed to be the preferred sites of star formation \citep{Andre2014}.  However, we still do not know if the magnetic field plays an important role in filament formation and evolution, as {\em Planck} could not spatially resolve filaments, which often have widths $\sim$0.1\,pc \citep{Arzoumanian2019}. 
Polarization observations of dense filaments and cores are particularly challenging as dust grains embedded in filaments are expected to be less well aligned, and thus polarization levels are expected to be lower \citep{Andersson2015}. 

\begin{figure}
    \centering
    \includegraphics[width=3.0in]{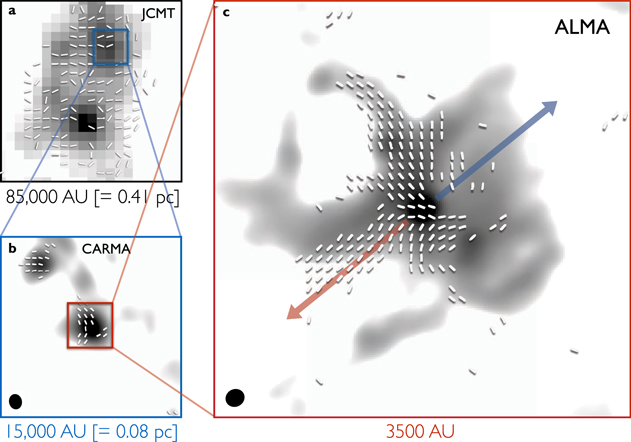}
    \includegraphics[width=3.0in]{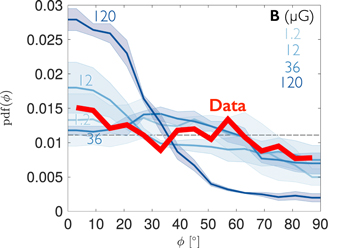}
    \caption{\textit{Figures from \cite{Hull2017a}. (left) A multi-scale polarimetric view towards the Ser-emb 8 protostar. (right) With increased statistics from increasingly sensitive polarimeters, statistics such as the histogram of relative orientations (HRO) can be used to compare models to measurements and in turn extract physical parameters from polarimetric observations.}}
    \label{fig:Hull}
\end{figure}

Multi-scale comparisons of the orientation between column density structures and the magnetic field, indicate that core scale fields may have magnetic energy density equal to or less than that of turbulence \citep[][ see Fig.~\ref{fig:Hull} (left)]{Hull2017a}. In contrast, in similar studies of cloud scale fields the magnetic energy appears to be at least as strong as turbulence \citep{PlanckXXXV,Soler2017}. Filaments and cores could therefore represent the densities at which magnetic energy becomes subdominant to gas kinetic energy \citep{CYChen2016}.

Systematic surveys of magnetic fields in filaments and cores are beginning, with instruments such as the TolTEC instrument on the LMT ($\lambda_{central}$ = 1.1, 1.4, and 2.1mm; best FWHM resolution of 5$^{\prime \prime}$;\citealt{WardThompson2017,Pattle2018}), POL-2/SCUBA2 on the JCMT ($\lambda_{central}$ = 450 and 850\,$\mu m$, with FWHM resolution of 14$^{\prime \prime}$ at 850\,$\mu m$), and the HAWC+ polarimeter on SOFIA \citep[50-214 $\mu$m, best FWHM resolution of 6$^{\prime \prime}$;][]{Harper2018}, all utilizing large focal plane cameras.  These instruments provide the ability to resolve filaments in nearby clouds, but are typically limited by sensitivity to observe only high column density filaments ($A_V\,>10$\,mag). A complete survey of fields within all filaments across many clouds, with resolution to isolate sub-filamentary structures seen in many nearby clouds (e.g.~\citealt{Hacar2018}) is needed.  Such a survey would probe whether magnetic fields provide significant support against gravitational collapse, and elucidate the role magnetic fields play in filament and core formation. With increased sensitivity, statistical quantities such as the histogram of relative orientations \citep[HRO; ][]{Soler2013} can be utilized to compare models of clouds, filaments, and cores with data (See Fig.~\ref{fig:Hull} (right)). 



Weak magnetic fields can still strongly inhibit protostellar disk formation if the field is parallel to the angular momentum axis of the disk, a conundrum known as the magnetic braking catastrophe \citep{Allen2003,WursterLi2018}.  The extent to which magnetic fields limit disk formation, perhaps due to misalignment between the angular momentum and magnetic field \citep{Joos2012}, is an important outstanding question. Large surveys of magnetic fields on filament and core scales with single dish polarimeters, combined observations of disks properties from ALMA and the EVLA, will be able to determine whether there is a correlation between disk sizes and masses and core/cloud magnetization and field geometry.
Polarimetry using interferometers such as ALMA or potentially the ngVLA can trace magnetic fields at even higher resolution, down to the scales of of protostellar envelopes and protoplanetary disks \citep{Hull2018b,HullZhang2019}.
\section{Conclusions}

To summarize, progress in understanding the magnetized diffuse ISM and the role of magnetic fields in formation of stars in molecular clouds requires:
\begin{itemize}
\item {\em High sensitivity polarization maps of the entire sky.}  All-sky surveys of dust polarization will allow detailed studies of magnetic fields and magnetized turbulence in the diffuse interstellar medium.  Since most of the magnetic field sampled with polarized dust emission above the Galactic plane is located within $\sim$200\,pc of the sun, high sensitivity surveys with $<$\,1.5$^{\prime}$ resolution will resolve scales below 0.1\,pc.  In addition, all-sky polarization maps from proposed experiments such as PICO will produce detailed magnetic field maps of thousands of molecular clouds.
\item {\em Detailed polarization maps with $\leq$10$^{\prime \prime}$ resolution over a large spatial dynamic range.}  To study the interaction between turbulence, magnetic fields, feedback, and gravity within molecular clouds, we require detailed polarization maps that can both resolve magnetic fields on $<<$0.1\,pc scales of cores and filaments, while also having the sensitivity to produce high-fidelity maps of the emission from low-column-density molecular cloud envelopes. Balloon-borne polarimeters will soon produce sub-arcmin resolution maps of many molecular clouds; however, higher sensitivity and resolution maps are needed from next-generation satellites such as OST \citep{OST2018} and SPICA \citep{Rodriguez2018}.
\item {\em Next-generation single-dish cameras with high resolution.} Polarimeters on large single-dish ground based telescopes, such as TolTEC on LMT \citep{toltec2018}, POL-2 on JCMT \citep{pol22018}, NIKA2pol on IRAM \citep{Adam2018}, and future instruments on proposed facilities like AtLAST \citep{Mroczkowski2018}, are needed to map magnetic field in dense structures, such as in cores and filaments.
\item {\em Continued development of polarization capabilities for interferometers.} Improved polarization sensitivity and resolution for ALMA and new facilities like the ngVLA \citep{ngVLA2018}, are need to survey magnetic fields in protostellar and protoplanetary disks.
\end{itemize}


\bibliography{ms}
\bibliographystyle{compact}

\end{document}